\documentclass[english]{ourlematema} 
\usepackage{graphicx}
\graphicspath{{images/}}

\usepackage{xcolor}
\DeclareMathOperator{\Gr}{Gr}

\title{From Feynman Diagrams to the Amplituhedron: \\ A Gentle Review}

\author{S. De}
\address{%
Brown University\\
Providence, USA\\
\email{shounak\_de@brown.edu}
}
\author{D. Pavlov}
\address{%
Max Planck Institute for Mathematics in the Sciences\\
Leipzig, Germany\\
and\\
Technische Universität Dresden\\
Dresden, Germany\\
\email{dmitrii.pavlov@mis.mpg.de}
}
\author{M. Spradlin}
\address{%
Brown University\\
Providence, USA\\
and\\
Brown Theoretical Physics Center\\
Providence, USA\\
\email{marcus\_spradlin@brown.edu}
}
\author{A. Volovich}
\address{%
Brown University\\
Providence, USA\\
\email{anastasia\_volovich@brown.edu}
}

\date{2024/10/13}

\begin{document}

\maketitle

\begin{abstract}

In this article we review, for a mathematical audience, the computation of (tree-level) scattering amplitudes in Yang-Mills theory in detail, in order to bridge the gap in understanding of the subject between mathematicians and physicists. In particular we demonstrate explicitly how the same formulas for six-particle NMHV helicity amplitudes are obtained from summing Feynman diagrams and from computing the canonical form of the $n=6, k=1, m=4$ amplituhedron.

\end{abstract}

\tableofcontents

\newpage

\section{Preliminaries on Amplitudes}
\label{sec:preliminaries}

We work in complexified Minkowski space-time $\mathbb{C}^{1,3}$ with the mostly minus metric, denoting the inner product of two vectors $a$, $b$ by
\begin{align}
\label{eq:innerproduct}
a \cdot b = a^0 b^0 - a^1 b^1 - a^2 b^2 - a^3 b^3\,.
\end{align}
\emph{Scattering amplitudes} (henceforth simply ``amplitudes'') of $n > 2$ massless vector particles, denoted generically by $A(p_i, \epsilon_i)$, are certain functions of $n$ \emph{momenta} $p_1,\ldots,p_n \in \mathbb{C}^{1,3}$ and $n$ \emph{polarizations} $\epsilon_1,\ldots,\epsilon_n \in \mathbb{C}^{1,3}$ subject to:
\begin{align}
\label{eq:c1}
1. & \ \text{each } p_i \ne 0,\\
\label{eq:c2}
2. & \ p_i \cdot p_i = \epsilon_i \cdot p_i = \epsilon_i \cdot \epsilon_i = 0 \text{ for all } i,\\
\label{eq:c3}
3. & \ \text{and \emph{momentum conservation}:} \sum_{i=1}^n p_i = 0\,.
\end{align}
Amplitudes must satisfy additional mathematical and physical constraints too numerous to review here, but three important general requirements are:
\begin{enumerate}
\item amplitudes must be linear in each polarization vector,
\begin{align}
A(p_i, z_i \epsilon_i) = A(p_i, \epsilon_i)\prod_{i=1}^n z_i \qquad \forall z_i \in \mathbb{C}\,,
\label{eq:linear}
\end{align}
\item invariant under \emph{gauge transformations}
\begin{align}
A(p_i, \epsilon_i + z_i p_i) = A(p_i, \epsilon_i) \qquad \forall z_i \in \mathbb{C}\,,
\label{eq:gaugeinvariance}
\end{align}
\item and \emph{Lorentz invariant}
\begin{align}
A(R p_i, R \epsilon_i) = A(p_i, \epsilon_i) \qquad \forall R \in SO(1,3,\mathbb{C})\,.
\end{align}
\end{enumerate}
The class of
amplitudes we focus on beginning in the next section manifest the last
property by depending only on the Lorentz invariant products $p_i \cdot p_j$, $\epsilon_i \cdot p_j$ and $\epsilon_i \cdot \epsilon_j$.

\section{Feynman Rules for Planar Yang-Mills Theory}
\label{sec:YMTheoryAmplitudes}

In this article, we focus on a particular class of amplitudes called the \emph{tree-level} amplitudes of \emph{planar Yang-Mills theory}. We write $A_n$ to denote the $n$-particle amplitude in this class, suppressing the implied dependence on the $p_i$, $\epsilon_i$ for notational efficiency. In this section, we explain diagrammatic \emph{Feynman rules}, which, in principle, can be used to compute $A_n$ for any $n$. More specifically, the rules we describe are known as \emph{color-ordered Feynman rules}, but we have no need to review the formalism that explains the meaning of this phrase and instead refer the reader to~\cite[Section 2.1]{Dixon:1996wi} for more details.

A \emph{Feynman diagram} for planar Yang-Mills theory is a connected, plane tree graph in which $n>2$ vertices (called \emph{external}) have degree 1, and the remaining vertices (called \emph{internal}) have degree 3 or 4, together with labeling of the external vertices $1,\ldots,n$ in clockwise order as one encircles the diagram from the outside. The $n$ edges connected to the external vertices are called \emph{external} edges, while all other edges are called \emph{internal}. The former naturally inherit the labeling $1,\ldots,n$ of the external vertices they are attached to, and we label the remaining edges $n+1, n+2, \ldots$ arbitrarily. To each Feynman diagram we associate a certain mathematical expression according to the following rules.

First, we assign an arbitrary direction to each internal edge and an outward pointing direction to each external edge. Next we assign a pair $p_e, \epsilon_e$ of elements of $\mathbb{C}^{1,3}$ to each edge $e$. For the $n$ external edges, these are the momenta and polarizations of the $n$ particles whose amplitude we are computing, which necessarily satisfy~(\ref{eq:c1}) and~(\ref{eq:c2}), but the momenta and polarizations assigned to internal edges are considered general elements of $\mathbb{C}^{1,3}$.

\begin{figure}[ht]
\begin{center}
    \includegraphics[width=0.50\textwidth]{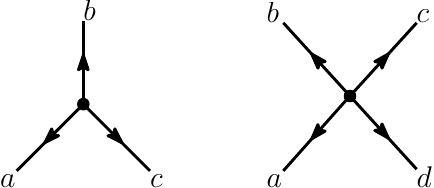}
\end{center}
\caption{Cubic and quartic vertices in Feynman diagrams are associated with the vertex factors~(\ref{eq:V3}) and~(\ref{eq:V4}) respectively.}
\label{fig:YMvertices}
\end{figure}

Next, we write down a product of vertex factors read off from this labeled diagram; the factor associated to each cubic or quartic vertex (see Fig.~\ref{fig:YMvertices}) is respectively:
\begin{align}
\label{eq:V3}
V_3(a,b,c) &= \frac{i}{\sqrt{2}} (\epsilon_a \cdot \epsilon_b)(\epsilon_c \cdot (p_a{-}p_b)) + {\rm cyclic}(a, b, c)\,, \\
V_4(a,b,c,d) &= i (\epsilon_a \cdot \epsilon_c)(\epsilon_b \cdot \epsilon_d) - \frac{i}{2} (\epsilon_a \cdot \epsilon_b) (\epsilon_c \cdot \epsilon_d) - \frac{i}{2} (\epsilon_b \cdot \epsilon_c) (\epsilon_d \cdot \epsilon_a)\,.
\label{eq:V4}
\end{align}
Here $i=\sqrt{-1}$. The term $\mathrm{cyclic}(a,b,c)$ in \eqref{eq:V3} is the instruction to sum over the cyclic shifts of the indices $(a,b,c)$. That is, we add the terms obtained by replacing $(a,b,c)$ with $(c,a,b)$ and $(b,c,a)$.
These formulas assume that all edges are directed outward; for every inward directed edge $a$, one should reverse the sign of $p_a$ and use $\epsilon_a^*$ instead of $\epsilon_a$ in these formulas (here $*$ denotes element-wise complex conjugation). We will account for this by using negative indices for ingoing edges, with the convention that $\epsilon_{-a} = \epsilon_a^*$ and $p_{-a} = - p_a$. Finally, we act on the product of vertex factors with the differential operator
\begin{align}
-i \prod_{e} \frac{-i}{p_e \cdot p_e} \frac{\partial}{\partial \epsilon_e} \cdot \frac{\partial}{\partial \epsilon_e^*}\,,
\end{align}
where the product runs over all internal edges $e$.

For example, the Feynman diagram
\begin{center}
    \includegraphics[scale=0.7]{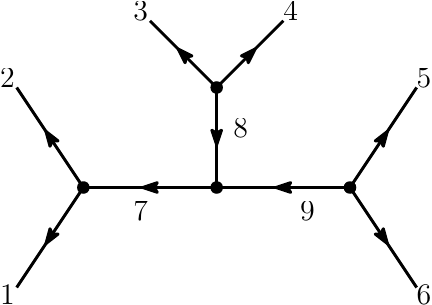}
\end{center}
evaluates to
\begin{align}
\label{eq:example2}
-i \left(\prod_{e=7}^9 \frac{-i}{p_e \cdot p_e} \frac{\partial}{\partial \epsilon_e} \cdot \frac{\partial}{\partial \epsilon_e^*}\right) V_3(1,2,-7) V_3(3,4,8) V_3(5,6,9) V_3(7,-8,-9)\,.
\end{align}

Note that the product of vertex factors is always manifestly linear in each $\epsilon_e$ and $\epsilon_e^*$. Therefore, the differential operators completely remove all dependence on the polarizations associated to the internal edges. The resulting expressions still apparently depend on the momenta associated to the internal edges, but these should be eliminated and are uniquely determined in terms of the external $p_i$ by demanding that the sum of incoming momenta equals the sum of outgoing momenta at each internal vertex. For example, in~(\ref{eq:example2}) we have
\begin{align}
p_7 = p_1 + p_2, \qquad p_8 = -p_3 - p_4, \qquad p_9 = -p_5 - p_6\,.
\end{align}
The result of this procedure is a quantity that depends only on the external momenta and polarizations.

\begin{figure}
\begin{center}
    \includegraphics[width=0.9\textwidth]{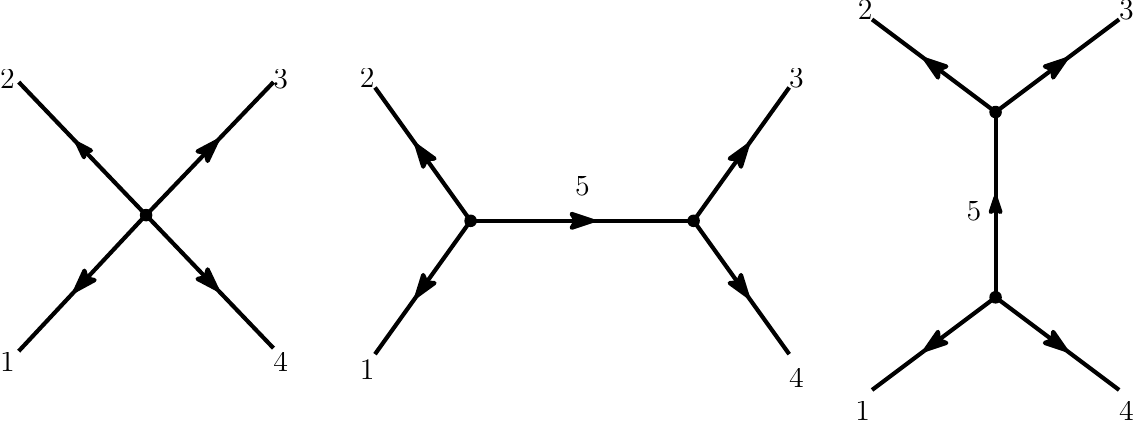}
\end{center}
\caption{The three Feynman diagrams contributing to the four-particle amplitude $A_4$ in planar Yang-Mills theory.}
\label{fig:fourpointexamplegraph}
\end{figure}

Let us work out and simplify one example in detail. Evaluating the second Feynman diagram in Fig.~\ref{fig:fourpointexamplegraph} gives
\begin{align}
\label{eq:example1}
-i  \frac{-i}{p_5 \cdot p_5} \frac{\partial}{\partial \epsilon_5} \cdot \frac{\partial}{\partial \epsilon_5^*} V_3(1,2,5) V_3(-5,3,4)\,.
\end{align}
Plugging in the expression for $V_3$ from~(\ref{eq:V3}) we obtain
\begin{equation}
\begin{split}
&= \frac{1}{2} \frac{1}{p_5 \cdot p_5} \left[ \epsilon_1 \cdot \epsilon_2(p_1{-}p_2) +
\epsilon_2 (\epsilon_1 \cdot (p_2{-}p_5)) + \epsilon_1 (\epsilon_2 \cdot (p_5{-}p_1)) \right]\\ 
& \qquad\qquad \cdot \left[ \epsilon_3 ( \epsilon_4 \cdot (- p_5{-}p_3)) + \epsilon_3 \cdot \epsilon_4 (p_3{-}p_4) + \epsilon_4 (\epsilon_3 \cdot (p_4{+}p_5))\right]\\
&=\frac{1}{2} \frac{1}{(p_1 + p_2)^2} \left[ \epsilon_1 \cdot \epsilon_2(p_1{-}p_2) +
2 \epsilon_2 (\epsilon_1 \cdot  p_2) -2 \epsilon_1 (\epsilon_2 \cdot p_1) \right]\\
&\qquad\qquad\qquad \cdot \left[ -2 \epsilon_3 ( \epsilon_4 \cdot  p_3) + \epsilon_3 \cdot \epsilon_4 (p_3{-}p_4) + 2 \epsilon_4 (\epsilon_3 \cdot  p_4)\right].
\end{split}
\end{equation}
To go from the first expression to the second we have plugged $p_5 = - p_1 - p_2$ into the first line and $p_5 = p_3 + p_4$ into the second, which eliminates some terms thanks to $\epsilon_i \cdot p_i = 0$.

Finally, the $n$-particle tree-level amplitude $A_n$ of planar Yang-Mills theory is obtained by summing the expressions associated by the above rules over all distinct Feynman diagrams with $n$ external vertices. For example, the four-particle amplitude is obtained from the three diagrams shown in Fig.~\ref{fig:fourpointexamplegraph}:
\begin{equation}
\begin{split}
A_4 &= (\epsilon_1 \cdot \epsilon_3) (\epsilon_2 \cdot \epsilon_4) - \frac{1}{2} (\epsilon_1 \cdot \epsilon_2) (\epsilon_3 \cdot \epsilon_4) - \frac{1}{2} (\epsilon_2 \cdot \epsilon_3)(\epsilon_4 \cdot \epsilon_1)\\
&+\frac{1}{2} \frac{1}{(p_1 + p_2)^2} \left[ \epsilon_1 \cdot \epsilon_2(p_1{-}p_2) +
2 \epsilon_2 (\epsilon_1 \cdot  p_2) -2 \epsilon_1 (\epsilon_2 \cdot p_1) \right]\\
&\qquad\qquad\qquad \cdot \left[ -2 \epsilon_3 ( \epsilon_4 \cdot  p_3) + \epsilon_3 \cdot \epsilon_4 (p_3{-}p_4) + 2 \epsilon_4 (\epsilon_3 \cdot  p_4)\right]\\
&+\frac{1}{2} \frac{1}{(p_2 + p_3)^2} \left[ \epsilon_2 \cdot \epsilon_3(p_2{-}p_3) +
2 \epsilon_3 (\epsilon_2 \cdot  p_3) -2 \epsilon_2 (\epsilon_3 \cdot p_2) \right]\\
&\qquad\qquad\qquad \cdot \left[ -2 \epsilon_4 ( \epsilon_1 \cdot  p_4) + \epsilon_4 \cdot \epsilon_1 (p_4{-}p_1) + 2 \epsilon_1 (\epsilon_4 \cdot  p_1)\right].
\label{eq:A4}
\end{split}
\end{equation}

\begin{figure}
\begin{center}
     \includegraphics[width=0.935\textwidth]{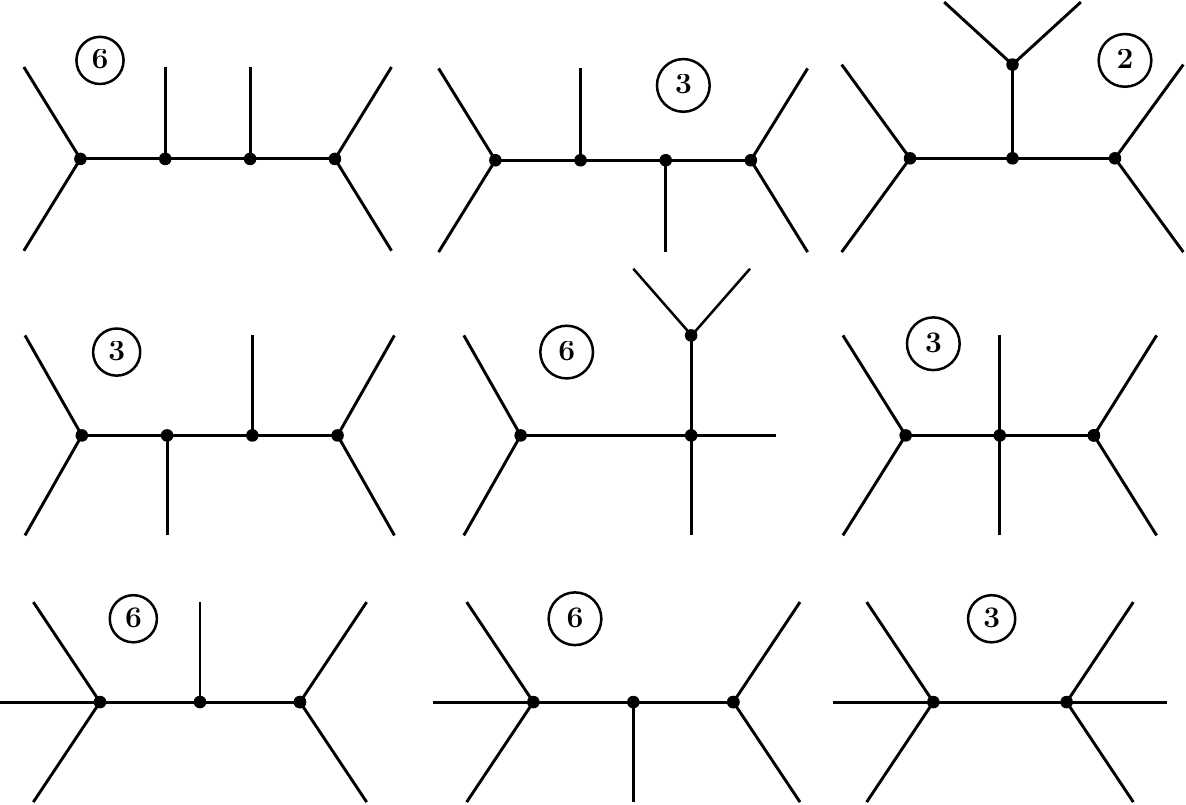}
\end{center}
\caption{The nine different graph topologies with 6 external vertices. To obtain a Feynman diagram one must assign a labeling $1,\ldots,6$ to the external vertices in clockwise order. The numbers in circles indicate the number of distinct labelings for each topology; altogether a total of 38 Feynman diagrams contribute to the amplitude $A_6$. If we had used non-color-ordered Feynman diagrams, one would have to sum over all permutations of labelings, not just cyclic permutations; in this case, there would be 220 contributing diagrams.}
\label{fig:sixpointgraphs}
\end{figure}

We end this section by tabulating also the three-particle amplitude, which has a single contributing Feynman diagram and is just
\begin{align}
\label{eq:A3}
A_3 = \frac{1}{\sqrt{2}} (\epsilon_1 \cdot \epsilon_2) ( \epsilon_3 \cdot (p_1{-}p_2)) + {\rm cyclic}(1,2,3)\,.
\end{align}
The calculation of the four-particle amplitude~(\ref{eq:A4}) already yields a rather complicated expression, and it is clear that the complexity grows considerably for higher $n$, not only because the number of diagrams increases, but also because intermediate expressions become very large. For $n=5, 6$ there are respectively 10, 38 Feynman diagrams (the latter are described in Fig.~\ref{fig:sixpointgraphs}) and it would be impractical to display explicit expressions for the corresponding amplitudes $A_n$. In the next section, we introduce a new set of variables that greatly simplifies the bookkeeping, thus simplifying these calculations while simultaneously revealing hidden mathematical structure that is completely obscured in the Feynman diagram formalism.

\section{Helicity Amplitudes}

A \emph{spinor} is a non-zero element of $\mathbb{C}^2$. \emph{Helicity amplitudes} of $n>2$ massless particles, introduced in~\cite{Xu:1986xb} and denoted generically by $A(\lambda_i, \Tilde{\lambda}_i, h_i)$, are certain functions of a choice of $n$ \emph{helicites} $h_i \in \frac{1}{2} \mathbb{Z}$ and $2n$ spinors $\lambda_i$ and $\Tilde{\lambda}_i$, which we think of respectively as $2\times 1$ matrices $\lambda_i = (\lambda_i^1\ \lambda_i^2)^T$ and $1\times 2$ matrices $\Tilde{\lambda}_i = (\Tilde{\lambda}_i^{1}\ \Tilde{\lambda}_i^{2})$, subject to
\begin{align}
\label{eq:momcon}
\sum_{i=1}^n \lambda_i \Tilde{\lambda}_i = 0\,.
\end{align}
Collectively we call the $\lambda_i$ and $\Tilde{\lambda}_i$ \emph{spinor helicity variables}. The role of helicities in our computations is that they pick particular polarization vectors $\epsilon_i^{h_i}$ to be plugged into the expressions coming from Feynman rules. The superscript $h_i$ is just a label, it has no meaning of an exponent. We refer the reader to~\cite[Chapter 2]{Elvang:2013cua} for a comprehensive treatment of helicity amplitudes, which must satisfy:
\begin{enumerate}
\item homogeneity in the spinor helicity variables
\begin{align}
\label{eq:c4}
A\left(z_i \lambda_i, z_i^{-1} \Tilde{\lambda}_i, h_i\right) = A(\lambda_i, \Tilde{\lambda}_i, h_i) \prod_{i=1}^n z_i^{-2 h_i} \qquad \forall z_i \in \mathbb{C}^*
\end{align}
\item and \emph{Lorentz invariance}, which, due to the isomorphism of Lie algebras ${\mathfrak{so}_{\mathbb{C}}(3,1)\cong \mathfrak{sl}_{\mathbb{C}}(2) \oplus \mathfrak{sl}_{\mathbb{C}}(2)}$, takes the form
\begin{align}
A(R \lambda_i,  \Tilde{\lambda_i} \Tilde{R}, h_i) = A(\lambda_i, \Tilde{\lambda}_i, h_i)\qquad  \forall R, \Tilde{R} \in SL(2,\mathbb{C})\,.
\end{align}
\end{enumerate}
By the First fundamental theorem of invariant theory for $SL(2, \mathbb{C})$ (see e.g. \cite[Theorem 3.2.1]{Sturmfels:2008it}), the latter implies that we can think of any helicity amplitude as a function of the Lorentz invariant quantities defined by
\begin{align}
    \langle i j \rangle = \det (\begin{matrix}
        \lambda_i & \lambda_j
    \end{matrix})
    \quad \text{and} \quad [ij] = \det(\begin{matrix}
        \Tilde{\lambda}_i^T & \Tilde{\lambda}_j^T
    \end{matrix}).
    \label{eq:shLIinnerprod}
\end{align}

Helicity amplitudes contain precisely the same information as the amplitudes we discussed in Sec.~\ref{sec:preliminaries}, just packaged in a slightly different way (for more details please refer to \cite{Dixon:1996wi,Elvang:2013cua}). Restricting our attention to massless vector particles, which have helicities $|h_i|= 1$, the correspondence is
\begin{align}
 A(p_i, \epsilon_i^{h_i}) = A(\lambda_i, \Tilde{\lambda}_i, h_i)
\end{align}
with the identification (henceforth we write $\epsilon^\pm$ as shorthand for $\epsilon^{\pm 1}$, where $\pm 1$ is the helicity label)
\begin{align}
\label{eq:pee}
p_i = \lambda_i \Tilde{\lambda}_i, \quad \epsilon^+_i = \sqrt{2} \frac{r_i \Tilde{\lambda}_i}{\langle ri \rangle}, \quad
\epsilon^-_i = - \sqrt{2} \frac{\lambda_i \Tilde{r}_i}{[ir]}\,.
\end{align}
Here $r_i$ and $\Tilde{r}_i$ are arbitrary spinors and we define
\begin{align}
\langle ir \rangle = \det\begin{pmatrix}\lambda_i & r_i \end{pmatrix},
\qquad
[ir] = \det\begin{pmatrix} \Tilde{\lambda}_i^T & \Tilde{r}_i^T\end{pmatrix}.
\end{align}
Now it remains only to explain that in~(\ref{eq:pee}), where each left-hand side is understood as element of $\mathbb{C}^{1,3}$ while each right-hand side is a $2 \times 2$ matrix, we adopt the identification
\begin{align}
\label{eq:mommat}
a = (a^0, a^1, a^2, a^3) \quad \leftrightarrow \quad
    \begin{pmatrix}
        a^0-a^3 & a^1 + ia^2\\
        a^1-ia^2 & a^0 + a^3
    \end{pmatrix}\,,
\end{align}
which has the property that $a \cdot a$ equals the determinant of the corresponding matrix. Note also that
\begin{align}
\label{eq:pdotp}
p_i \cdot p_j = \frac{1}{2} \langle ij\rangle [ij]\,.
\end{align}

Let us comment on the identification~(\ref{eq:pee}). First, the dependence on the arbitrary spinors drops out because any change in $r_i$ or $\Tilde{r}_i$ only shifts each polarization $\epsilon_i$ by a multiple of the corresponding $p_i$, which has no effect on an amplitude due to the gauge invariance property~(\ref{eq:c2}). Second, if we rescale $\lambda_i \to t_i \lambda_i$ and $\Tilde{\lambda}_i \to t_i^{-1} \Tilde{\lambda}_i$ for any $t_i \in \mathbb{C}^*$, then the $p_i$ are unchanged, but $\epsilon_i^{h_i}$ scales as $t_i^{-2h_i}$; this explains the connection between the linearity property~(\ref{eq:c2}) of amplitudes and the scaling~(\ref{eq:c4}) property of helicity amplitudes. The choice of numerical constants in~(\ref{eq:pee}) provides a convenient normalization choice that simplifies some formulas, specifically $\epsilon_i^+ \cdot \epsilon_i^- = - 1$.

Let us work out some examples in order to illustrate this new notation. For simplicity we choose all of the reference spinors $r_i \equiv r$ and $\Tilde{r}_i \equiv \Tilde{r}$ to be the same. In this case the following formulae hold for the remaining products of momenta and polarization vectors:
\begin{align} \label{eq:pe}
    \epsilon_i^+ \cdot \epsilon_j^- = -\dfrac{\langle jr \rangle[ir]}{\langle i r\rangle [j r]}\,, \qquad
    p_i\cdot \epsilon_j^+ = \dfrac{\langle ir \rangle [j i]}{\sqrt{2}\langle jr\rangle}\,,
    \qquad p_i\cdot \epsilon_j^- =  \dfrac{[ ir]\langle ji\rangle}{\sqrt{2}[ jr]}\,.
\end{align}

\subsection{Three-particle amplitudes}

Three-particle helicity amplitudes are special because the spinor helicity variables are highly constrained. Specifically, note that
\begin{align}
\langle 12 \rangle [12] = 2 p_1 \cdot p_2 = (p_1 + p_2)^2 = p_3^2 = 0\,,
\end{align}
where we have used~(\ref{eq:pdotp}) as well as $p_1 + p_2 + p_3 = 0$ and $p_1^2=p_2^2=p_3^2 = 0$. It follows that we must have either $\langle 12 \rangle = 0$ or $[12] = 0$. Repeating this calculation for other pairs of indices leads to the conclusion that in fact we must have
\begin{align}
\text{ either all } \langle ij\rangle = 0 \text{ or all } [ij] = 0, \text{ if } n = 3\,.
\end{align}

It is straightforward, if a little tedious, to check that if all $\langle ij \rangle = 0$, then the only non-zero three-particle helicity amplitudes are those for which two particles have helicity $+1$ and the third has helicity $-1$. Similarly, if all $[ij] = 0$ then the only non-zero helicity amplitudes are those for which two particles have helicity $-1$ and the third has helicity $+1$.

As an example let us work out the helicity amplitude $A(1^-,2^-,3^+)$ in detail, in the case when all $[ij] = 0$. Consequently from~(\ref{eq:pe}) we see that $p_i \cdot \epsilon_j^+ = 0$ for all $i,j$. From~(\ref{eq:A3}) we then have initially only two non-zero terms
\begin{align}
     \sqrt{2} A(1^-, 2^-, 3^+) =
     (\epsilon_2^- \cdot \epsilon_3^+) ( \epsilon_1^- \cdot (p_2{-}p_3)) +
     (\epsilon_3^+ \cdot \epsilon_1^-) ( \epsilon_2^- \cdot (p_3{-}p_1))\,.
\end{align}
We use momentum conservation $p_3 = -p_1 - p_2$ and the condition $p_1\cdot \epsilon_1^- = 0$ to rewrite the first term as $2 (\epsilon_2^- \cdot \epsilon_3^+)(\epsilon_1^- \cdot p_2)$. A similar calculation for the second term leads to
\begin{align}
     A(1^-, 2^-, 3^+) &=
     \sqrt{2} (\epsilon_2^- \cdot \epsilon_3^+) ( \epsilon_1^- \cdot p_2 ) -
     \sqrt{2} (\epsilon_3^+ \cdot \epsilon_1^-) ( \epsilon_2^- \cdot p_1) \nonumber \\
     &= -\dfrac{\langle 2r \rangle[3r]}{\langle 3 r\rangle [2 r]} \dfrac{[ 2r]\langle 12\rangle}{[ 1r]}
     + \dfrac{\langle 1r \rangle[3r]}{\langle 3 r\rangle [1 r]}\dfrac{[ 1r]\langle 21\rangle}{[ 2r]} \nonumber \\
     &= -\frac{\langle 12 \rangle [3r] (  \langle 2r \rangle [2r] + \langle 1r \rangle [1r] ) }{[1r] [2r] \langle 3r \rangle} \nonumber \\
     &= \frac{\langle 12 \rangle [3r]^2 }{[1r] [2r] }~,
\end{align}
where in the last line we used momentum conservation in the form
\begin{align}
\langle 1r \rangle [1r] + \langle 2r \rangle [2r] + \langle 3r \rangle [3r] = 2 p_r \cdot p_1 +
2 p_r \cdot p_2 + 2 p_r \cdot p_3 = 0\,.
\end{align}
Here $p_r = r \Tilde{r}$ is the momentum corresponding to the reference spinors $r$ and $\Tilde{r}$. Another similar application of momentum conservation allows us to write
\begin{align}
    \dfrac{[3r]}{[1r]}  = \dfrac{\langle 12 \rangle}{\langle 23\rangle} \quad \text{and} \quad \dfrac{[3r]}{[2r]} = \dfrac{\langle 12\rangle}{\langle 31\rangle}
\end{align}
which leads to the final expression
\begin{align}
    A(1^-,2^-,3^+) = \frac{\langle12\rangle^3}{\langle23\rangle\langle31\rangle}\,.
\label{eq:3pointMHV}
\end{align}

When all $\langle ij \rangle = 0$ we similarly have for example
\begin{align}
\label{eq:3pointMHVbar}
A(1^+,2^+,3^-) = \frac{[ 12 ]^3}{ [ 23 ] [31 ] }\,.
\end{align}
The results~(\ref{eq:3pointMHV}) and~(\ref{eq:3pointMHVbar}) exemplify a  general \emph{parity symmetry} for helicity amplitudes under which the $\lambda_i$ and $\Tilde{\lambda}_i$ spinor variables are exchanged while flipping the sign of each helicity:
\begin{align}
\label{eq:paritydef}
A(1^{h_1}, 2^{h_2}, \ldots) = A(1^{-h_1}, 2^{-h_2}, \ldots) \rvert_{\langle ij \rangle \leftrightarrow [ij]}\,.
\end{align}

Other three-particle helicity amplitudes can be computed from~(\ref{eq:A3}) with similar effort; we summarize here the results~\cite{Benincasa:2007xk}
\begin{align}
A(1^{h_1}, 2^{h_2}, 3^{h_3}) = \begin{cases}
\langle 12 \rangle^{h_3-h_1-h_2} \langle 23 \rangle^{h_1-h_2-h_3} \langle 31 \rangle^{h_2-h_1-h_3} & \text{if } h = -1,\\
[12]^{h_3-h_1-h_2}[23]^{h_1-h_2-h_3}[31]^{h_2-h_1-h_3} & \text{if } h= +1, \\
0 & \text{otherwise}\,,
\end{cases}
\label{eq:allthreepoint}
\end{align}
where $h=h_1+h_2+h_3$ is the total helicity.

\subsection{MHV helicity amplitudes}

The calculation of three-particle helicity amplitudes outlined in the previous section is already rather tedious, though the final result~(\ref{eq:allthreepoint}) is strikingly compact. Needless to say, the complexity grows considerably for higher $n$, but some simple cases remain.  Specifically, for $n > 3$, it turns out that all helicity amplitudes with total helicity $h = \sum_i h_i$ satisfying $|h| \ge n - 2$ vanish. The simplest non-vanishing helicity amplitudes are those with $h=n-4$, which are called \emph{MHV amplitudes} and are given by the \emph{Parke-Taylor} formula~\cite{Parke:1986gb}
\begin{align}
A(1^+,\ldots, i^-, \ldots, j^-, \ldots, n^+) = \frac{\langle ij\rangle^4}{\langle 12 \rangle \langle 23 \rangle \cdots \langle n1 \rangle}
\label{eq:parketaylor}\,.
\end{align}
Helicity amplitudes with $h=4-n$ are called $\overline{\text{MHV}}$ amplitudes and are given by the parity conjugate of this formula. The formula~(\ref{eq:parketaylor}) and the claims about helicity amplitudes with $|h| \ge n-2$ vanishing can be compared against the sum of Feynman diagrams for relatively small values of $n$ with the assistance of a computer algebra system. We provide code to check this for all helicity amplitudes with $n \le 6$ \cite{mathrepo}.

\subsection{Six-Particle NMHV amplitudes}\label{sec:sixpointNMHV}

Helicity amplitudes with total helicity $h = n - 4 - 2 k$ are called \emph{N${}^k$MHV amplitudes}. The smallest value of $n$ for which there are helicity amplitudes that are neither MHV nor $\overline{\text{MHV}}$ is $n=6$. Here there are three independent types of NMHV amplitudes. In order to display formulae for these we introduce the notation
\begin{align}
s_{ijk} = [ij]\langle ij\rangle + [ik] \langle ik\rangle + [jk] \langle jk \rangle\,, \qquad
[i|P_{jk}|\ell\rangle = [ij]\langle j\ell\rangle + [ik] \langle k \ell \rangle\,.
\end{align}
Then we have
\begin{multline}
    A(1^-,2^-,3^-,4^+,5^+,6^+) =
     \frac{[4|P_{32}|1\rangle^3}{[23] [34] [2|P_{34}|5\rangle \langle 56 \rangle \langle 61 \rangle s_{234}}\\ + \frac{[6|P_{54}|3\rangle^3}{[61] [12] [2|P_{34}|5\rangle \langle 34 \rangle \langle 45 \rangle s_{612}}\,.
    \label{eq:splithelicitysh}
\end{multline}
\begin{multline}
    A(1^+,2^-,3^+,4^-,5^+,6^-) =\\
    \frac{[15]^4 \langle 24\rangle^4}{[56] [61] [1|P_{23}|4\rangle [5|P_{43}|2\rangle \langle 23 \rangle \langle 34 \rangle s_{234}}+ {\rm cyc}_2 + {\rm cyc}_4\,,
\end{multline}
where the notation indicates a sum of three terms, with the first term as shown and the second and third obtained from the first by replacing all indices $i \to i + 2~{\rm mod}~6$ and $i \to i + 4~{\rm mod}~6$, respectively.
\begin{multline}
    A(1^+,2^+,3^-,4^+,5^-,6^-) =
    \frac{[4|P_{12}|3\rangle^4}{[45][56] [4|P_{32}|1\rangle [6|P_{12}|3\rangle \langle 12 \rangle \langle 23 \rangle s_{123}}\\
     +\frac{[24]^4 \langle 56\rangle^3}{[23] [34] [2|P_{34}|5\rangle [4|P_{32}|1\rangle \langle 61 \rangle s_{234}} +
    \frac{[12]^3 \langle 35 \rangle^4}{[61] [2|P_{34}|5\rangle [6|P_{54}|3\rangle \langle 34 \rangle \langle 45 \rangle s_{345}}\,.
\end{multline}
All other six-particle NMHV amplitudes can be obtained from these three by some cyclic rotation of the indices possibly accompanied by a parity conjugation using~(\ref{eq:paritydef}). Equivalent formulas for these helicity amplitudes were first obtained in~\cite{Mangano:1987xk,Berends:1987cv}, and the ones we have presented were first given in~\cite{Roiban:2004ix,Britto:2004ap}. We consider it essentially impossible to check these formulas by hand starting from Feynman diagrams. Instead, we provide \texttt{Mathematica} code to perform this verification \cite{mathrepo}. However it is relatively straightforward to derive these formulas analytically from the BCFW recursion relations \cite{Britto:2005fq}. The connection between the BCFW recursion and the amplituhedron was mathematically worked out in \cite{Even:2025} and \cite{Even:2023}. 

\section{From Spinor Helicity Variables to Momentum Twistors}

In this section we introduce momentum twistors, which provide a convenient parameterization of the spinor helicity variables for configurations of $n$ particles and are the natural variables for discussing the amplituhedron.

\subsection{Definition of momentum twistors}

We define a \emph{configuration of $n \ge 4$ particles} to be a collection of $n$ spinor helicity variables $\{ \lambda_i, \Tilde{\lambda}_i\}$ satisfying $\langle i\ i{+}1\rangle \ne 0$ and $[i\ i{+}1] \ne 0$ for all $i$ (here and in all that follows, indices are always understood mod $n$), as well as momentum conservation~\eqref{eq:momcon}. A \emph{twistor} is a point in $\mathbb{CP}^3$ and we denote the homogeneous coordinates of a twistor $Z$ as $(Z^1 : Z^2 : Z^3 : Z^4)$. The words \emph{momentum twistor} emphasize the use of twistors for parameterizing the momenta of particles in a scattering amplitude. Momentum twistors were introduced by Hodges~\cite{Hodges:2009hk}, who proved the following result. 

\begin{thm}\label{thm:twistors}
Let $\{ \lambda_i, \Tilde{\lambda}_i\}$ be a configuration of $n \ge 4$ particles. Then there exist $n$ twistors $Z_i = (\lambda_i^1:\lambda_i^2:\mu_i^1:\mu_i^2)\in\mathbb{CP}^3$ and $n$ dual twistors $\Tilde{Z}_{i} = (\Tilde{\mu}_{i}^{1}:\Tilde{\mu}_{i}^{2}:\Tilde{\lambda}_{i}^{1}:\Tilde{\lambda}_{i}^{2})\in(\mathbb{CP}^3)^*$ satisfying the following conditions for each $i$:
\begin{enumerate}
    \item For all $i$, the plane defined by $\Tilde{Z}_i$ contains $Z_{i-1}$, $Z_{i}$, $Z_{i+1}$ and
    \item For all $i$, the points $Z_i$, $Z_{i+1}$, $Z_{i+2}$ and $Z_{i+3}$ are not contained in a plane. 
\end{enumerate}
\end{thm}

In particular, one sees that a spinor $\lambda_i$ can be ``extended'' to a twistor $Z_i$ by adding two $\mu$-coordinates, and the same is true for $\Tilde{\lambda}_i$ and $\Tilde{Z}_i$. Explicit in the presentation of this theorem is a particular choice of coordinates on projective space such that the spinor helicity brackets may be recovered from the twistors and dual twistors via
\begin{align}
\label{eq:allij}
\langle i j \rangle = Z_i^T I Z_j, \qquad [ij] = \Tilde{Z}_i^T \Tilde{I} \Tilde{Z}_j
\end{align}
using
\begin{align}
    I= \begin{pmatrix}
        0 & 1 & 0 & 0\\
        -1 & 0 & 0 & 0\\
        0 & 0 & 0 & 0\\
        0 & 0 & 0 & 0
    \end{pmatrix} \quad \text{and} \quad \Tilde{I} = \begin{pmatrix}
        0 & 0 & 0 & 0\\
        0 & 0 & 0 & 0\\
        0 & 0 & 0 & 1\\
        0 & 0 & -1 & 0
    \end{pmatrix}.
\end{align}
The relation between primal and dual twistors in Theorem~\ref{thm:twistors} is given by \cite[Eqn. (6)]{Hodges:2009hk}:
\begin{align}\label{eq:allimp}
    \Tilde{Z}_{i}^{A} = \dfrac{\epsilon^{ABCD}Z_{i-1}^B Z_i^C Z_{i+1}^D}{\langle i{-}1 \ i\rangle \langle i \ i{+}1 \rangle}\,,
\end{align}
This formula simply gives the coordinates of the plane $\Tilde{Z}_i$ containing the points $Z_{i-1}$, $Z_i$, $Z_{i+1}$. Here and further in the article, we use the Einstein summation convention that repeated indices (in this case, the indices $B$, $C$ and $D$ on the right-hand side) are summed over their natural range (in this case, 1 through 4) without writing the summation explicitly, and $\epsilon^{ABCD} = \epsilon_{ABCD}$ is the Levi-Civita symbol defined as the signature of $(A,B,C,D)$ if $(A,B,C,D)$ is a permutation of $(1,2,3,4)$, and 0 otherwise. Finally we define the coordinates
\begin{align}
\langle i\,j\,k\,l \rangle = \det(Z_i \, Z_j \, Z_k \, Z_l ) = \epsilon_{ABCD} Z_i^A Z_j^B Z_k^C Z_l^D \,.
\label{eq:twistorcoordinates}
\end{align}

Of most immediate practical use to us is the following converse, which is a consequence of~\eqref{eq:allimp} and the formula~\eqref{eq:squaretwistor} proven below.
\begin{thm}
Let $\{Z_i\}$ be a collection of $n \ge 4$ points in $\mathbb{CP}^3$ that satisfies $\langle i\,i{+}1 \rangle \ne 0$ and $\langle i\,i{+}1\,i{+}2\,i{+}3 \rangle\ne 0$ for all $i$. Then for $A=1,2$
\begin{align}
\label{eq:hodgesconversea}
\lambda_i^A &= Z_i^A, \\
\Tilde{\lambda}_i^{A} &= \frac{\langle i{-}1 \ i \rangle Z_{i+1}^{A+2} + \langle i{+}1 \ i{-}1 \rangle Z_{i}^{A+2} + \langle i \ i{+}1 \rangle Z_{i-1}^{A+2}}{\langle i{-}1 \ i \rangle \langle i \ i{+}1 \rangle}
    \label{eq:hodgesconverseb}
\end{align}
is the corresponding configuration of $n$ particles.
\end{thm}

\subsection{A warm-up calculation with momentum twistors}

Equation~\eqref{eq:allimp} implies for example that
\begin{align} \label{eq:squaretwistor}
    [i\ i+1] = \dfrac{\langle i-1\ i\ i+1\ i+2\rangle}{\langle i-1\ i\rangle \langle i\ i+1\rangle \langle i+1\ i+2\rangle}.
\end{align}
Here we derive this formula in detail in order to give a better idea of how to manipulate spinor helicity and momentum twistor variables. We begin by noting that
\begin{align}
\tilde{I}^{AB} = \frac{1}{2} \epsilon^{ABCD} I^{CD}\,,
\end{align}
which we can plug into~\eqref{eq:allij} and then use~\eqref{eq:allimp} to write
\begin{align} \label{eq:iinext}
    [i\, i{+}1] = \Tilde{I}^{AB} \Tilde{Z}_{i}^A\Tilde{Z}_{i+1}^{B} = \dfrac{1}{2}\epsilon^{ABCD} I^{CD}\dfrac{\epsilon_{AEFG}Z^E_{i-1}Z^F_{i}Z^G_{i+1} \epsilon_{BHJK}Z_i^H Z_{i+1}^J Z^K_{i+2}}{\langle i{-}1\, i\rangle \langle i\, i{+}1\rangle^2 \langle i{+}1\, i{+}2\rangle}.
\end{align}
The first indices in the Levi-Civita symbols are contracted using the identity
\begin{multline} \label{eq:deltas}
    \epsilon^{ABCD} \epsilon_{AEFG} =\\ \delta_E^B \delta_F^C \delta_G^D + \delta_E^C \delta_F^D \delta_G^B + \delta_E^D \delta_F^B \delta_G^C - \delta_E^B \delta_F^D \delta_G^C  - \delta_E^D \delta_F^C \delta_G^B -
    \delta_E^C \delta_F^D \delta_G^B,
\end{multline}
where $\delta$ is the Kronecker delta symbol: $\delta^X_Y = 1$ if $X=Y$ and is zero otherwise. This identity simply states that $\epsilon^{ABCD} \epsilon_{AEFG}$ is equal to $1$ is $(BCD)$ is an even permutation of $(EFG)$ and is equal to $-1$ if $(BCD)$ is an odd permutation of $(EFG)$ (and is $0$ otherwise).

When we plug \eqref{eq:deltas} into \eqref{eq:iinext}, the second term gets contracted with $\epsilon_{BHJK}$ to give $\delta_E^C \delta_F^D \epsilon_{GHJK}$. This quantity vanishes when further contacted with $Z_{i+1}^G Z_{i+1}^J$ due to antisymmetry in $G \leftrightarrow J$. For the same reason, the third, fifth, and sixth terms also do not contribute to $[i\,i{+}1]$. Due to the antisymmetry of $I^{CD}$, the first and the fourth terms give the same contribution to $[i\,i{+}1]$, canceling the factor of $\frac{1}{2}$. Thus, we are left with just
\begin{align}
    [i\,i{+}1] = \dfrac{\delta_E^B \delta_F^C \delta_G^D \epsilon_{BHJK} \, I^{CD} \, Z^E_{i-1}Z^F_i Z^G_{i+1} Z^H_i Z^J_{i+1} Z^K_{i+2}}{\langle i{-}1\,i \rangle \langle i\,i{+}1 \rangle^2 \langle i{+}1\,i{+}2 \rangle}~,
\end{align}
which, with the help of~\eqref{eq:allij} and~\eqref{eq:twistorcoordinates}, simplifies to
\begin{align}
    \dfrac{(\epsilon_{EHJK}Z_{i-1}^E Z_i^H Z_{i+1}^J Z_{i+2}^K)(I^{FG}Z_i^F Z_{i+1}^G)}{\langle i{-}1\,i\rangle \langle i\,i{+}1\rangle^2 \langle i{+}1\,i{+}2\rangle} = \dfrac{\langle i{-}1\,i\,i{+}1\,i{+}2\rangle}{\langle i{-}1\,i\rangle \langle i\,i{+}1\rangle \langle i{+}1\,i{+}2\rangle},
\end{align}
completing the proof of~\eqref{eq:squaretwistor}.

\subsection{A six-particle NMHV amplitude in momentum twistors}

We are now equipped to end this section by rewriting the helicity amplitude (\ref{eq:splithelicitysh}) in terms of the momentum twistors using the relations defined above. Using similar manipulations to the one carried out in the previous subsection, one can show that
\begin{align}
    [4|P_{56}|1\rangle = \frac{\langle1345\rangle}{\langle34\rangle \langle45\rangle}~~,~~[6|P_{12}|3\rangle = \frac{\langle1356\rangle}{\langle56\rangle \langle61\rangle}~~,~~[2|P_{34}|5\rangle = \frac{\langle1235\rangle}{\langle12\rangle \langle23\rangle}
\end{align}
along with
\begin{align}
    s_{612} = \frac{\langle2356\rangle}{\langle23\rangle\langle56\rangle}\,, \quad s_{234} = \frac{\langle1245\rangle}{\langle12\rangle\langle45\rangle}\,.
\end{align}
Putting everything together, we conclude that the helicity amplitude~(\ref{eq:splithelicitysh}) can be expressed in momentum twistor variables as
\begin{multline}
    A_6(1^-,2^-,3^-,4^+,5^+,6^+) = \frac{\langle12\rangle^4 \langle23\rangle^4}{\langle12\rangle\langle23\rangle\langle34\rangle\langle45\rangle\langle56\rangle\langle61\rangle}  \\
    \times \frac{1}{\langle1235\rangle}\left(\frac{\langle1356\rangle^3}{\langle2356\rangle \langle1236\rangle \langle1256\rangle} + \frac{\langle1345\rangle^3}{\langle1234\rangle \langle1245\rangle \langle2345\rangle} \right).
\end{multline}
The other helicity amplitudes displayed in Sec.~\ref{sec:sixpointNMHV} are considerably more messy when expressed in momentum twistor variables. Instead of presenting these formulas, we now introduce a vastly more efficient bookkeeping mechanism that naturally combines all NMHV amplitudes into a single object.

\section{Supervariables and Superamplitudes}

Superamplitudes provide a convenient bookkeeping device for assembling all $\binom{n}{k+2}$ $n$-particle N${}^k$MHV helicity amplitudes into a single object.

In this section, we introduce Grassmann partners for the spinor helicity variables and momentum twistors, then write down formulas for MHV superamplitudes and the six-particle NMHV superamplitude, and finally explain how to verify that these superamplitudes encapsulate all of the helicity amplitudes discussed in previous sections.

\subsection{Supervariables} \label{sec:supervars}

First, we introduce some basic properties of \emph{Grassmann variables} (also called anti-commuting). If $\theta_a$ is a collection of Grassmann variables indexed by some set with elements $a$, then $\theta_a \theta_b = - \theta_b \theta_a$ and integration over the $\theta$'s is defined in the sense of Berezin by
\begin{align}
\int d \theta_a = 0\, \qquad \int \theta_a \, d\theta_a = 1
\label{eq:berezin}
\end{align}
and extended by linearity.

Next, we introduce Grassmann partners of the momentum twistor variables. A \emph{momentum supertwistor} is a point in the projective space $\mathbb{CP}^{3|4}$. This is $\mathbb{CP}^7$ in which we treat the first four homogeneous coordinates differently from the last four. Namely, we write the homogeneous (super) coordinates of a momentum supertwistor as $(Z^1: Z^2: Z^3: Z^4:\chi^1: \chi^2: \chi^3:\chi^4)$ where $Z^A$ are ordinary bosonic (i.e., commuting) coordinates and the $\chi^A$ are Grassmann variables. The straightforward extension of Theorem \ref{thm:twistors} to $\mathbb{CP}^{3|4}$ demonstrates that a collection of $n \ge 4$ momentum supertwistors satisfying $\langle i\,i{+}1 \rangle \ne 0$ and $\langle i\,i{+}1\,i{+}2 \,i{+}3\rangle \ne 0$ for all $i$ provides, via~\eqref{eq:hodgesconversea} and~\eqref{eq:hodgesconverseb}, together with the analogous formula
\begin{align}
    \eta_i^A &= \frac{\langle i{-}1 \, i \rangle \chi_{i+1}^A + \langle i{+}1 \, i{-}1 \rangle \chi_{i}^A + \langle i \, i{+}1 \rangle \chi_{i-1}^A}{\langle i{-}1 \, i \rangle \langle i \, i{+}1 \rangle}\,,
    \label{eq:incidence4}
\end{align}
a configuration of $n$ particles, supplemented with an associated collection of Grassmann variables $\eta_i$ that also satisfy \emph{supermomentum conservation}:
\begin{align}
\label{eq:smcon}
    \sum_{i=1}^n \lambda_i \eta_i = 0\,.
\end{align}
For future convenience we also introduce the notation
\begin{align}
\overline{\eta}_i = \prod_{A=1}^4 \eta_i^A\,, \qquad d\overline{\eta}_i = \prod_{A=1}^4 d \eta_i^A,\, \qquad
d\overline{\eta} = \prod_{i=1}^n d\overline{\eta}_i\,.
\end{align}

\subsection{MHV superamplitudes}

The $n$-particle \emph{MHV superamplitude} is the homogeneous polynomial of degree 8 in the Grassmann variables given by
\begin{align}
\mathcal{A}^{\rm MHV}_n = \frac{\delta^{(8)}(q)}{\langle 1 2 \rangle \langle 23 \rangle \cdots \langle n1 \rangle}\,,
\end{align}
where $q = \sum_i \lambda_i \eta_i$ and
\begin{align} \label{eq:grassmanndelta}
\delta^{(8)}(q) = \prod_{A=1}^4 \sum_{i<j} \langle ij \rangle \eta_i^A \eta_j^A\,.
\end{align}
Upon expanding out the product, it is evident that the coefficient of the term $\overline{\eta}_i \overline{\eta}_j$ in $\mathcal{A}_n^{\rm MHV}$ is precisely the Parke-Taylor formula~\eqref{eq:parketaylor} for the $n$-particle helicity amplitude in which particles $i, j$ have helicity $-1$ and the remaining particles have helicity $+1$. Using the properties of Berezin integration \eqref{eq:berezin}, this fact may also be expressed as
\begin{align}
    A(1^+,\ldots, i^-, \ldots, j^-, \ldots, n^+) &= \mathcal{A}_n^{\rm MHV}\bigg|_{\text{coefficient of}~ \overline{\eta}_i \overline{\eta}_j} \nonumber \\
    &= \int  \left(\prod_{l \neq i,j} \overline{\eta}_l\right) \mathcal{A}^{\rm MHV}_n\,  d\overline{\eta} \label{eq:mhvextract}\,, 
\end{align}
where we have utilized the expressions \eqref{eq:berezin} and \eqref{eq:grassmanndelta} to obtain 
\begin{align}
\label{eq:deltaq}
     \delta^{(8)}(q)\bigg|_{\text{coefficient of}~ \overline{\eta}_i \overline{\eta}_j} = \int \left(\prod_{l \neq i,j} \overline{\eta}_l\right) \delta^{(8)}(q) \,d\overline{\eta} = \langle ij\rangle^4\,. 
\end{align}

\subsection{The six-particle NMHV superamplitude}

The $n$-particle \emph{NMHV superamplitude} $\mathcal{A}^{\rm NMHV}_n$ is a polynomial of degree 12 in the Grassmann variables that encapsulates NMHV helicity amplitudes in a manner analogous to~\eqref{eq:mhvextract}. If, for notational simplicity, we let $A^{(n)}_{ijk}$ denote the NMHV helicity amplitude where particles $i,j,k$ have helicity $-1$ and the rest have helicity $+1$, then
\begin{align}
\label{eq:nmhvextract}
A^{(n)}_{ijk} = \int  \left(\prod_{l \neq i,j,k} \overline{\eta}_l\right) \mathcal{A}^{\rm NMHV}_n\, d\overline{\eta}
= \mathcal{A}^{\rm NMHV}_n \bigg|_{\text{coefficient of}~ \overline{\eta}_i \overline{\eta}_j \overline{\eta}_k}\,.
\end{align}

We now claim that the six-particle NMHV superamplitude is given by~\cite[Eqn. (31)]{Hodges:2009hk}
\begin{align} \label{eq:nmhvsuper}
    \mathcal{A}_6^{\text{NMHV}} = \mathcal{A}_6^{\text{MHV}} \cdot \left( [61234] + [61245] + [62345]\right)
\end{align}
in terms of the 5-bracket
\begin{align}
\label{eq:fivebracket}
[abcde] = \frac{\prod_{A=1}^4 \left( \langle abcd \rangle \chi_e^A + {\rm cyclic} \right)}{\langle abcd \rangle \langle bcde \rangle \langle cdea \rangle \langle deab \rangle \langle eabc \rangle},
\end{align}
where ``$+$ cyclic'' is the instruction to sum cyclically over the labels $\{a,b,c,d,e\}$ in a fashion similar to the cyclic product structure in the denominator, and $\chi$ is a collection of Grassmann variables introduced in Section \ref{sec:supervars}. The quantity
\begin{align}
\label{eq:ratiofunction}
{\mathcal{P}}_6^{\rm NMHV}(Z_i, \chi_i) = [61234] + [61245] + [62345]
\end{align}
appearing in~\eqref{eq:nmhvsuper} is called the 6-particle \emph{NMHV ratio function}. In Section \ref{sec:6} we will see that the amplituhedron computes precisely this ratio function. 

Finally we describe how to explicitly verify that the simple expression~\eqref{eq:nmhvsuper} indeed encapsulates all $\binom{6}{3} = 20$ six-particle NMHV helicity amplitudes tabulated in Sec.~\ref{sec:sixpointNMHV}. First plugging~\eqref{eq:nmhvsuper} into~\eqref{eq:nmhvextract} gives
\begin{align}
A_{ijk}^{(6)} = \int  \left(\prod_{l \neq i,j,k} \overline{\eta}_l\right)  \frac{\delta^{(8)}(q)}{\langle 1 2 \rangle \langle 23 \rangle \cdots \langle 61 \rangle} \left([61234] + [61245] + [62345]\right)  d \overline{\eta}\,.
\end{align}
Then we use the Grassmann delta function $\delta^8(q)$ from \eqref{eq:grassmanndelta} to carry out the integration over the $\overline{\eta}_i$ and $\overline{\eta}_j$ Grassmann variables using~\eqref{eq:deltaq}, which leads to
\begin{align}
A_{ijk}^{(6)}
&= \frac{\langle ij \rangle^4}{\langle 12 \rangle \langle 23 \rangle \cdots \langle n1 \rangle}
\left([61234] + [61245] + [62345]\right) \rvert_{S_{ij}}
\bigg|_{\text{coefficient of}~ \overline{\eta}_k}
\end{align}
where $\rvert_{S_{ij}}$ is an instruction to first solve~\eqref{eq:smcon} to express $\eta_i$ and $\eta_j$ in terms of the other four $\eta$'s, and then rewrite the $\chi$'s appearing in the 5-brackets in terms of the $\eta$'s by inverting~\eqref{eq:incidence4}. Because of the complexity of the intermediate expressions involved, this step is best carried out with the help of a computer algebra system. We provide code to perform this verification for all 20 NMHV helicity amplitudes \cite{mathrepo}.

\section{Superamplitudes from the Amplituhedron} \label{sec:6}

In this section, we explain how to extract the NMHV ratio function~\eqref{eq:ratiofunction} from the amplituhedron $\mathcal{A}_{6,1,4}(\mathcal{Z})$. Our treatment closely follows Sec.~4 of~\cite{Arkani-Hamed:2010wgm}. The observation that NMHV amplitudes are related to the volumes of simplices is due to Hodges~\cite{Hodges:2009hk} and inspired Arkani-Hamed and Trnka to introduce the amplituhedron as a suitable generalization for N${}^k$MHV amplitudes with $k>1$ in~\cite{Arkani-Hamed:2013jha}.

The \emph{positive Grassmannian} $\Gr_{k,n}^{\ge 0}$ \cite{Postnikov:2004} is the subset of the real Grassmannian $\Gr_{k,n}$ on which all Pl\"ucker coordinates are non-negative. For positive integers $n,k,m$ such that $k+m\le n$ and $n>3$, the (tree-level) \emph{amplituhedron} $\mathcal{A}_{n,k,m}(\mathcal{Z})$ is the image of the positive Grassmannian $\Gr_{k,n}^{\ge 0}$ under the map
\begin{align}
[C] \to [C \mathcal{Z}] \in \Gr_{k,k+m}
\end{align}
induced by a totally positive $n \times (k+m)$ matrix $\mathcal{Z}$. Here $[C]$ refers to an element of $\Gr_{k,n}$ represented by a positive $k \times n$ matrix $C$.

It is conjectured that $\mathcal{A}_{n,k,m}(\mathcal{Z})$ is a \emph{positive geometry} \cite[Conjecture 6.3]{Arkani-Hamed:2017pg}, which among other things means that there exists a top-dimensional \emph{canonical form} $\Omega_{n,k,m}(Y;\mathcal{Z})$ that has no singularities inside $\mathcal{A}_{n,k,m}(\mathcal{Z})$ and logarithmic singularities on its boundary. Here $Y$ represents an element of $\Gr_{k,k+m}$ and the notation serves to emphasize that we are considering a family of forms \emph{on} $\Gr_{k,k+m}$ \emph{indexed} by a choice of $n \times (k+m)$ matrix $\mathcal{Z}$. The main conjecture of Arkani-Hamed and Trnka, who introduced the amplituhedron in~\cite{Arkani-Hamed:2013jha}, is that $\Omega_{n,k,4}(Y;\mathcal{Z})$ encodes the $n$-particle N${}^k$MHV superamplitude in a manner that we now describe.

Let us restrict to the case $n=6, k=1$, $m=4$ that is the focus of this paper. In this case we are interested in the form $\Omega_{6,1,4}(Y;\mathcal{Z})$, which necessarily can be written in the form
\begin{align}
\Omega_{6,1,4}(Y;\mathcal{Z}) = \omega_{6,1,4}(Y;\mathcal{Z}) \mu_{\mathbb{P}^4}(Y)\,,
\end{align}
where
\begin{align}
\mu_{\mathbb{P}^4}(Y)=
\sum_{a=1}^5 (-1)^a Y_a dY_1\wedge\ldots\wedge \widehat{d Y_a}\wedge \ldots\wedge dY_5
\end{align}
is the standard meromorphic top-form on $\Gr_{1,5} \cong \mathbb{P}^4$. Then according to the proposal of~\cite{Arkani-Hamed:2013jha}, the recipe that relates the canonical form on the amplituhedron to the NMHV ratio function~\eqref{eq:ratiofunction} is
\begin{align}
\label{eq:dictionary}
\mathcal{P}^{\rm NMHV}_6(Z_i, \chi_i)  = \int  \omega_{6,1,4}(Y_0;\mathcal{Z}) \, d^4 \phi
\end{align}
where $Y_0 = (0:0:0:0:1)$, $\mathcal{Z}$ is the $6 \times 5$ matrix whose $i$-th row is
\begin{align}
\label{eq:curlyz}
\mathcal{Z}_i = (Z_i^1, Z_i^2, Z_i^3, Z_i^4, \phi^A \chi_{i}^A)\,,
\end{align}
$\phi^A$ is a collection of four Grassmann variables, with the integral in~\eqref{eq:dictionary} understood in the usual Berezinian sense. 

For the case under study, the amplituhedron $\mathcal{A}_{6,1,4}(\mathcal{Z})$ is a cyclic polytope in $\mathbb{P}^4$. Its vertices are given by the rows of the matrix $\mathcal{Z}$. This polytope admits a triangulation into three simplices $\Delta_{\{61234\}}$, $\Delta_{\{61245\}}$ and $\Delta_{\{62345\}}$, where $\Delta_{\{abcde\}}$ is the convex hull of $\{\mathcal{Z}_a, \mathcal{Z}_b, \mathcal{Z}_c, \mathcal{Z}_d, \mathcal{Z}_e\}$.

The canonical form of the amplituhedron is then the sum of the canonical forms of these three simplices:
\begin{align}
\label{eq:threeterms}
\omega_{6,1,4}(Y;\mathcal{Z}) = \omega_{\{61234\}}(Y; \mathcal{Z}) + \omega_{\{61245\}}(Y;\mathcal{Z})
+ \omega_{\{62345\}}(Y;\mathcal{Z})
\end{align}
where the canonical form of each simplex is
\begin{align}
\label{eq:simplex}
    \omega_{\{abcde\}}(Y;\mathcal{Z})=
    \dfrac{\langle abcde \rangle^4}{\langle Y a b c d \rangle \langle Y b c d e \rangle \langle Y c d e a \rangle \langle Y d e a b \rangle \langle Y e a b c \rangle}\,.
\end{align}
The denominator in this expression is the product of the facet equations of the simplex, and the factor in the numerator ensures that its residues at every vertex are $\pm 1$. Angular brackets in this expression denote the determinant.

When we set $Y = Y_0$ in~\eqref{eq:simplex} it is clear that each denominator factor $\langle Y abcd \rangle$ becomes simply the corresponding coordinate $\langle abcd \rangle$. Therefore
\begin{align}
\int  \omega_{\{abcde\}}(Y_0;\mathcal{Z})\, d^4 \phi &=
\frac{1}{\langle  a b c d \rangle \langle  b c d e \rangle \langle c d e a \rangle \langle  d e a b \rangle \langle e a b c \rangle} \int \langle abcde \rangle^4 \, d^4 \phi\nonumber \\
&= [abcde]
\end{align}
where in the second line the rules of Berezinian integration land us directly on~\eqref{eq:fivebracket}. Plugging in the sum of three terms in~\eqref{eq:threeterms} confirms the agreement with the NMHV ratio function~\eqref{eq:ratiofunction} via~\eqref{eq:dictionary}.

We note that $k=1$ amplituhedra are cyclic polytopes for any $m$ and $n$ \cite{Sturmfels:1988}, and the procedure of extracting the $n$-point NHMV amplitude from $\mathcal{A}_{n,1,4}(\mathcal{Z})$ is analogous to that described above. For the general case of the $n$-point $\text{N}^k\text{MHV}$ amplitude, we refer the reader to Section 7 of the original paper \cite{Arkani-Hamed:2013jha}.

We conclude by noting that computing amplitudes with the amplituhderon is conceptually simpler than by using Feynman diagrams. This is already visible in the $k=1$ case: the number of terms in a triangulation of the cyclic polytope on $n$ vertices is quadratic in $n$, while the number of color-ordered Feynman diagrams grows exponentially with $n$. 

\section*{Acknowledgements}

We are grateful to B.~Sturmfels for encouraging us to write this article and to M.~Parisi and J.~Trnka for valuable input. This work was supported in part by the US Department of Energy under contract DE-SC0010010 Task F (SD, MS, AV), and by Simons Investigator Award \#376208 (AV). 


\begin{thebibliography}{10}
\setlength{\itemsep}{-0.1mm}

\bibitem{Arkani-Hamed:2017pg}
N.~Arkani-Hamed, Y.~Bao and T.~Lam,
``Positive Geometries and Canonical Forms,''
JHEP \textbf{11}, 039 (2017)
[arXiv:1703.04541 [hep-th]].

\bibitem{Arkani-Hamed:2010wgm}
N.~Arkani-Hamed, J.~L.~Bourjaily, F.~Cachazo, A.~Hodges and J.~Trnka,
``A Note on Polytopes for Scattering Amplitudes,''
JHEP \textbf{04}, 081 (2012)
[arXiv:1012.6030 [hep-th]].

\bibitem{Arkani-Hamed:2013jha}
N.~Arkani-Hamed and J.~Trnka,
``The Amplituhedron,''
JHEP \textbf{10}, 030 (2014)
[arXiv:1312.2007 [hep-th]].

\bibitem{Benincasa:2007xk}
P.~Benincasa and F.~Cachazo,
``Consistency Conditions on the S-Matrix of Massless Particles,''
[arXiv:0705.4305 [hep-th]].

\bibitem{Berends:1987cv}
F.~A.~Berends and W.~Giele,
``The Six Gluon Process as an Example of Weyl-Van Der Waerden Spinor Calculus,''
Nucl. Phys. B \textbf{294}, 700-732 (1987).

\bibitem{Britto:2004ap}
R.~Britto, F.~Cachazo and B.~Feng,
``New recursion relations for tree amplitudes of gluons,''
Nucl. Phys. B \textbf{715}, 499-522 (2005)
[arXiv:hep-th/0412308 [hep-th]].

\bibitem{Britto:2005fq}
R.~Britto, F.~Cachazo, B.~Feng and E.~Witten,
``Direct proof of tree-level recursion relation in Yang-Mills theory,''
Phys. Rev. Lett. \textbf{94}, 181602 (2005)
[arXiv:hep-th/0501052 [hep-th]].

\bibitem{mathrepo}
S.~De, D.~Pavlov, M.~Spradlin and A.~Volovich. Auxiliary code. [\url{https://mathrepo.mis.mpg.de/FeynmanToAmplituhedron}], 2024. 

\bibitem{Dixon:1996wi}
L.~J.~Dixon,
``Calculating scattering amplitudes efficiently,''
Lectures at TASI 1995,
[arXiv:hep-ph/9601359 [hep-ph]].

\bibitem{Elvang:2013cua}
H.~Elvang and Y.-t.~Huang,
``Scattering Amplitudes,''
[arXiv:1308.1697 [hep-th]].

\bibitem{Even:2023}
C.~Even-Zohar, T.~Lakrec, M.~Parisi, R.~Tessler, M.~Sherman-Bennett and L.~Williams,
``Cluster algebras and tilings for the $m=4$ amplituhedron,'' 
[arXiv:2310.17727 [math]].

\bibitem{Even:2025}
C.~Even-Zohar, T.~Lakrec and R.~J.~Tessler, ``The amplituhedron BCFW triangulation,'' Invent. Math. \textbf{239}: 1-130 (2025)
[arXiv:2112.02703 [math-ph]].

\bibitem{Hodges:2009hk}
A.~Hodges,
``Eliminating spurious poles from gauge-theoretic amplitudes,''
JHEP \textbf{05}, 135 (2013)
[arXiv:0905.1473 [hep-th]].

\bibitem{Mangano:1987xk}
M.~L.~Mangano, S.~J.~Parke and Z.~Xu,
``Duality and Multi-Gluon Scattering,''
Nucl. Phys. B \textbf{298}, 653-672 (1988).

\bibitem{Parke:1986gb}
S.~J.~Parke and T.~R.~Taylor,
``An Amplitude for $n$ Gluon Scattering,''
Phys. Rev. Lett. \textbf{56}, 2459 (1986).

\bibitem{Postnikov:2004}
A.~Postnikov, ``Total positivity, Grassmannians, and networks,''
[arXiv:math/0609764
 [math]].

\bibitem{Roiban:2004ix}
R.~Roiban, M.~Spradlin and A.~Volovich,
``Dissolving $\mathcal{N}=4$ loop amplitudes into QCD tree amplitudes,''
Phys. Rev. Lett. \textbf{94}, 102002 (2005)
[arXiv:hep-th/0412265 [hep-th]].

\bibitem{Sturmfels:2008it}
B.~Sturmfels, ``Algorithms in Invariant Theory,'' 
Springer Science \& Business Media (2008).

\bibitem{Sturmfels:1988}
B.~Sturmfels, ``Totally positive matrices and cyclic polytopes,''
Linear Algebra Appl. \textbf{107}, 275-281 (1988).

\bibitem{Xu:1986xb}
Z.~Xu, D.~H.~Zhang and L.~Chang,
``Helicity Amplitudes for Multiple Bremsstrahlung in Massless Nonabelian Gauge Theories,''
Nucl. Phys. B \textbf{291}, 392-428 (1987).

\end{thebibliography}

\begin{small}

\end{small}

\end{document}